\documentclass[a4paper,fleqn,usenatbib]{mnras}

\usepackage[T1]{fontenc}
\usepackage{ae,aecompl}

\usepackage{graphicx}	
\usepackage{amsmath}	
\usepackage{amssymb}	
\usepackage{multirow}
\usepackage{hyperref}
\usepackage[flushleft]{threeparttable}
\usepackage[justification=centering]{caption}

\newcommand{\degree}{^{\circ}}

\title{OH maser toward IRAS~06056+2131: polarization parameters and evolution status}

\author[M. Darwish et al.]{
Darwish, M. S.$^{1,2}$,\thanks{E-mail: darwish.msk@gmail.com}
Richards, A. M. S.$^{3}$,
Etoka, S.$^{3}$,
Edris, K. A.$^{4}$,
\newauthor
Saad, S. M.$^{1,2}$,
Beheary, M. M.$^{4}$,
Fuller, G. A.$^{3}$
\\
$^{1}$Astronomy Department, National Research Institute of Astronomy and Geophysics (NRIAG), 11421 Helwan, Cairo, Egypt.\\
$^{2}$Kottamia Center of Scientific Excellence in Astronomy and Space Science (KCScE, STDF No. 5217, ASRT), Cairo, Egypt.\\
$^{3}$Jodrell Bank Centre for Astrophysics, Department of Physics $\&$ Astronomy, The University of Manchester, M13 9PL, UK.\\
$^{4}$Astronomy and Meteorology Department, Faculty of Science, Al-Azhar University, Cairo, Egypt.\\
}
\date{Accepted 2020 September 11. Received 2020 September 9; in original form 2020 July 22}

\pubyear{2020}

\begin{document}
\label{firstpage}
\pagerange{\pageref{firstpage}--\pageref{lastpage}}
\maketitle
%
\begin{abstract}
{We present high angular resolution observations of OH maser emission towards the high-mass star forming region IRAS 06056+2131. The observations were carried out using the UK radio interferometer array, Multi-Element Radio Linked Interferometer Network (MERLIN) in the OH main lines at 1665- and 1667-MHz, in addition to the OH satellite line at 1720-MHz. The results of this study reveal the small upper limit to the size of emission in the 1665-MHz line with an estimated total intensity of $\sim 4$ Jy. We did not detect any emission from the 1667-MHz and 1720-MHz lines. The full polarization mode of MERLIN enables us to investigate the magnetic field in the OH maser region. In this transition, a Zeeman pair is identified from which a magnetic strength of $\sim -1.5$ mG is inferred. Our results show that IRAS 06056+2131 is highly polarized, with $\sim$ 96 $\%$ circular polarization and $\sim$ 6 $\%$ linear polarization. The linear polarization angle is $\sim 29$$^{\circ}$, implying a magnetic field which could be aligned with the outflow direction detected toward this region, but the actual magnetic field direction has an uncertainty of up to 110$^{\circ}$ due to the possible effects of Faraday rotation. The star forming evolutionary status of the embedded proto-stellar object is discussed.}

\end{abstract}

%
\section{Introduction}\label{sec:intro}
Despite the important role that the high-mass stars play in the formation of stars and galaxies as well as the evolution of the universe, many questions are still open in this issue.
Different theoretical approaches have been introduced in order to answer the question of how massive stars form. One approach, proposed by \cite{2003ApJ...585..850M}, is known as the core accretion model. They suggested that the formation of massive stars is similar to that of low mass stars, where the dominant force is the magnetic field and sufficient mass accretion only occurs when the magnetic support is removed through jets and outflows \citep[e.g.][]{1979ApJ...230..204M,2006ApJ...646.1043M,2011ApJ...742L...9C,2013ApJ...779...96T,2017MNRAS.465.2254K}.

On the other hand, other authors \citep[e.g.][]{2002ApJ...576..870P,2004RvMP...76..125M,2005prpl.conf.8555K,2011MNRAS.414.2511V} argue that dynamical influences such as turbulence play more effective roles than the magnetic field, particularly in the early stages of massive star formation. Therefore, understanding the role of the magnetic field during the formation of massive stars can lead us to a better understanding of how such stars are formed.

Maser emission lines provide exceptionally high resolution probes to measure the small scale magnetic field strength and structure within 10s -- 1000s AU of high mass protostars \citep[e.g.][]{2007ApJ...670.1159F,2010MNRAS.404..134V,2013A&A...556A..73S,2017A&A...597A..43G,2019FrASS...6...66C}. Masers are also used to investigate the kinematics as well as the physical conditions surrounding massive protostellar objects from the onset of formation \citep[e.g.][]{1997MNRAS.289..203C,2000prpl.conf..327S,2004A&A...414..235S,2010MNRAS.406.1487B,2020MNRAS.493.4442D}. Hydroxyl (OH), water (H$_{2}$O) and methanol (CH$_{3}$OH) masers
are commonly used to investigate the kinematics as well as the physical conditions surrounding massive protostellar objects in the early stage(s) of their formation. Due to its paramagnetic nature, the OH radical is consider to be more sensitive than CH$_{3}$OH and H$_{2}$O for measuring the Zeeman effect directly and consequently measuring the magnetic fields strength toward these objects \citep[e.g.][]{1985cgd..conf..223C,2005A&A...434..213E,2007MNRAS.382..770G,10.1111/j.1365-2966.2012.21722.x,2007IAUS..242...37V,2012MNRAS.423..647E,2017A&A...608A..80E}. However, H$_{2}$O and CH$_{3}$OH are also important particularly in studying the morphology of the magnetic field since their linear polarization vectors are less affected by Faraday rotation than OH masers \citep{2006A&A...448..597V,2011A&A...527A..48S,2011A&A...533A..47S,2017ApJ...834..168M}.
Zeeman splitting of OH maser lines can provide the 3D orientation of magnetic fields towards massive protostellar objects. The compactness and brightness of masers allow polarization observations at high angular resolution using interferometers, such as e-MERLIN, the VLBA (Very Long Baseline Array) and the EVN (European VLBI Network).

{OH masers, particularly the main lines at 1665 and 1667 MHz, are known to be frequently associated with different evolutionary stages of high-mass star forming regions (HMSFRs) \citep[e.g.][]{2007IAUS..242..213E}. \cite{2011MNRAS.415.3872C} reported that OH masers are associated with Ultra-Compact H\,{\small II} (UC\,H\,{\small II}) regions \citep[see also,][]{1988MNRAS.231..205C,1990A&A...236..479B,2010MNRAS.401.2219B}, other authors \citep[e.g.][]{2010MNRAS.406.1487B,2007A&A...465..865E,1999PASP..111.1049G} found that OH masers can also be associated with H{\small II} regions, which represent a more advanced stage in the massive star formation time scale.}

{IRAS 06056+2131 (also known as AFGL6366S) is a high mass star-forming region located in the Gemini OB1 cloud, which is known to be one of the most massive molecular cloud complexes in the outer Galaxy. It is named after the IRAS source located at RA (2000)= 06$^h$ 08$^m$ 40$^s$.9, Dec (2000)= 21$\degree$ 31$^\prime$ $00''$  with an error ellipse of 30$\times$5 arcsec at a position angle (PA) of 91$^{\circ}$ \citep{joint1988infrared}, giving
 an uncertainty of $2\fs15$, $5\farcs0$ in RA and Dec, respectively.}

{This source  is one of a large sample of candidate high-mass young stellar objects (YSO) which were identified by \cite{1991A&A...246..249P}. The sample of 260 IRAS sources was divided into two subsamples based on their IRAS color. The first subsample, so-called ``high", is composed of sources where [25$\mu$m -12$\mu$m] > 0.57\footnote[1]{[$\lambda_{2} - \lambda_{1}$] is defined as $\log_{10}[F_{\lambda 2}/F_{\lambda 1}]$, where $F_{\lambda i}$ is the IRAS flux density in wavelength band $\lambda_{i} $.}, which fulfils \cite{1989ApJ...340..265W} criteria for objects associated with ultra-compact H\,{\small II} (UC\,H\,{\small II}) regions. The second subsample, so-called ``low", is composed of sources with [60$\mu$m-12$\mu$m] > 1.3, where different evolutionary stages can be found, extending from the stage prior to the UC\,H\,{\small II} detection to evolved sources \citep{1996A&A...308..573M}. According to \cite{1991A&A...246..249P}, IRAS 06056+2131 belongs to the ``high" subsample.}
{The estimated far-infrared luminosity of IRAS 06056+2131 is 5.83$\times$10$^3$ ${L_\odot}$ \citep[and references therein]{2018ApJS..235...31Y} while the distance to the source is thought to be in the range of 0.8 to 2.5 kpc \citep{1989A&A...221..295K,1994ApJS...91..659K}. We adopt a distance of 1.5 kpc as an average \citep{2005ApJ...625..864Z}. A CO (J=2-1) bipolar outflow was detected toward  IRAS 06056+2131 by \cite{1988ApJ...325..853S,2004A&A...426..503W} and \cite{2005ApJ...625..864Z}. \cite{1994ApJS...91..659K}, using the VLA at a resolution  $\leq$ 1$^\prime$ detected weak radio continuum  flux (less than 1 mJy) at 3.6 cm, within $8\farcs0$ of the IRAS source, while \cite{2010ApJS..188..123R}, through the Bolocam Galactic Plane Survey (BGPS), indicated that IRAS 06056+2131 is associated with mm continuum emission.}

{IRAS 06056+2131 is known to be associated with several maser species. A Class II 6.7-GHz methanol maser was first detected toward the source at 12.7 Jy and found to be offset from the nominal IRAS position by $7\farcs6$  \citep[see,][]{1995MNRAS.272...96C,2000A&AS..143..269S,2009A&A...507.1117X,2010A&A...517A..56F}. Water maser emission at 22-GHz was detected by \cite{1989A&A...221..295K} and \cite{2007PASJ...59.1185S} with peak flux densities of 23 Jy and 2.43 Jy, respectively, although \cite{1991A&A...246..249P} failed to detect any emission in this transition, suggesting that the water masers toward this source are variable. The first detection of the OH maser toward IRAS 06056+2131 was reported by \cite{1979A&AS...37....1T}, while the first positional association was made by \cite{1988MNRAS.231..205C}. The LSR (Local Standard of Rest) velocity for IRAS 06056+2131 given in SIMBAD is $2.5\pm1.6$ km s$^{-1}$, measured using SEST (Swedish-ESO Submillimeter Telescope) observations of CS by
\cite{1996A&AS..115...81B}  but there may be multiple sources within their $50''$ beam.}

{In this study, we aim to estimate the accurate position of the OH maser within the IRAS 06056+2131 region and contribute to a better understanding of the evolutionary status of the IRAS source. Additionally, we investigate the magnetic field in the maser region using full-polarization MERLIN observations. In Section \ref{sec:obser}, the observations and data reduction are described. The data analysis and results of the imaging are given in Section \ref{sec:analys}. Our discussion and conclusion are presented in Section \ref{sec:discu} and \ref{sec:concl}, respectively.}

\section{Observations and data reduction}
\label{sec:obser}
{The observations of the source IRAS 06056+2131 were carried out using the MERLIN interferometer. Observations of the OH maser emission were performed in full-polarization mode during February 2007. IRAS 06056+2131 was observed for 3 full tracks, switching between 1665~MHz and 1667~MHz in addition to 1720 MHz transitions about every 30 minutes during the observations.}

 {The target was observed in bands centred on 1665.402, 1667.357 and 1720.520~MHz, adjusted to fixed velocity with respect to the LSR in the correlator assuming a target velocity of 10~km~s$^{-1}$ \citep{2007A&A...465..865E}. The compact quasar 0617+210 was observed as a phase reference calibrator with position RA = 06$^h$ 20$^m$ 19$^s$.529, Dec = 21$\degree$ 02$^\prime$ $29\farcs501$. This was observed alternately with the target in an approximately 10 minute cycle, giving a total of about 6 hr useful data on-target at each frequency. 0617+210  was observed in the full 16 MHz bandwidth (`wide', 13 MHz useful) with two tunings, one covering 1665 and 1667 MHz and the other covering 1720 MHz.}

 {The data were extracted from the MERLIN archive and converted to FITS at Jodrell Bank Center for Astrophysics (JBCA) using local software (dprogs) \citep{Diamond03} and the Astronomical Image
Processing System (AIPS) software package (\url{http://www.aips.nrao.edu/cook.html}). In order to calibrate and reduce the data we have used the Common Astronomy Software Application (CASA) software package version 5.1.6 \citep{2007ASPC..376..127M}, the full documentation can be found at \url{https://casa.nrao.edu/casadocs}. The observational parameters for IRAS 06056+2131 are listed in Table~\ref{tab:MERLIN_data}. The observations and data reduction followed normal MERLIN procedures, see the MERLIN User Guide \citep{Diamond03}.}

{The OH maser lines were observed in a spectral bandwidth of 0.5 MHz (`narrow') corresponding to 80~km~s$^{-1}$ useful velocity range with a channel separation of 0.18~km~s$^{-1}$. The source 3C84 was observed as a bandpass calibrator in both wide and narrow configurations. At this frequency the flux density for 3C84  was set to be 17.05 Jy based on previous scaling using 3C286, which has a flux density around 13.6 Jy \cite{1977A&A....61...99B}, allowing for the resolution of MERLIN}.

{We used 3C84 to correct for the wide-narrow phase offset. The polarization leakage for each antenna was estimated using the un-polarized source 3C84 while the calibration and correction for the polarization position angle was carried out using the source 3C286, which has a known polarization angle of 33$^{\circ}$ in the image plane. These corrections, along with the bandpass table and the phase reference solutions for phase and amplitude, were applied to the target.}

\begin{figure}
\centering
\includegraphics [width=1.0 \columnwidth]{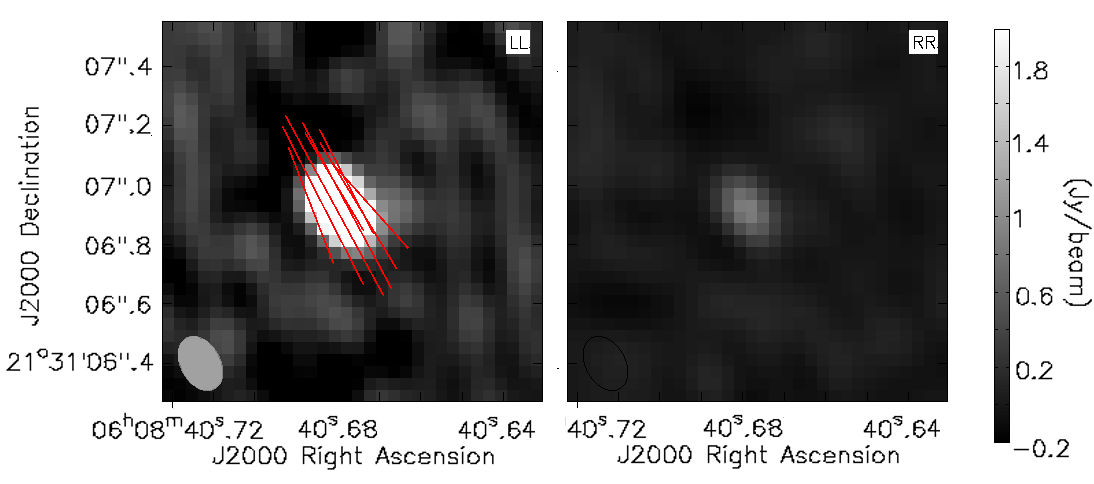}
     \caption{Maps of the clean image of the brightest target channel ($V_{\mathrm{LSR}}$= 10 km/s). The left panel shows left-hand circular polarization, while the right panel shows right-hand circular polarization. The ellipse in the left corner represents the MERLIN primary beam shape. The intensity is shown by the grey scale while the red lines represent the linear polarization vectors direction.}
  \label{fig:LL_RR_polvec}
\end{figure}

{The CLEAN algorithm in CASA package ``tclean" was used to clean and de-convolve the image cubes, using the default Cotton-Schwab based deconvolution \citep{1984AJ.....89.1076S} and the Hogbom minor cycle deconvolver \citep{1974A&AS...15..417H}. We cleaned all Stokes parameters $I$, $Q$, $U$ and $V$ (total intensity, linear and circular products, see Section \ref{sec:analys},  Equations 1-4), with the same mask, and similarly cleaned the correlator circular polarization products RR and LL.} The $\sigma_{\mathrm{rms}}$ noise in quiet channels was 0.03 Jy in $I$, $Q$, $U$ and $V$, and 0.04 Jy in RR and LL. The synthesised beam size was $0\farcs19$ $\times$ $0\farcs13$ at PA = 29.39$^{\circ}$.
{We made a linearly-polarized intensity (POLI) image using 0.03 Jy for de-biasing, and a polarization angle (POLA) cube, using a cut-off of 3 times the POLI $\sigma_{\mathrm{rms}}$ of 0.04 Jy (see Sect. \ref{sec:analys} , Eqs. 5 and 6)}.
Figure \ref{fig:LL_RR_polvec} displays the brightest channels of IRAS 06056+2131 after cleaning each circular polarization separately.

{In this work we use `component' to refer to a single patch of emission
 in a single channel (sometimes referred to as `spots'), and `feature' to refer
 to a series of components which are likely to make up a physical association.}
{We used the CASA task ``imfit" to determine the positions of the maser components(`spots') by fitting 2D Gaussian components to each patch of emission above a 3$\sigma_{\mathrm{rms}}$ threshold in each channel map for the total intensity (Stokes $I$) cube.}
{The OH maser position errors due to the angular separation and
the cycle time between the phase-reference source and the target, are
about 0\farcs008 and 0\farcs022, respectively. The stochastic error due to noise
for the peak is 0\farcs007. {Allowing for the phase
reference and telescope position errors, which are comparatively very small, we obtain a total astrometric error of 0\farcs025 in each direction}.}

{The position error for fitting emission imaged using a sparse array such as MERLIN is approximately (synthesised beam)/(signal-to-noise ratio) so the components were required to appear in at least three consecutive channels with positions within 0\farcs05 of their occurrence in each consecutive channel, such groups forming spectral features. It was apparent that the resulting components all occurred at the same position within the errors (see Section \ref{sec:analys}), in the $V_{\mathrm{LSR}}$ range 8.95 -- 10.53~km~s$^{-1}$}.

\begin{table*}
\begin{center}
\caption{IRAS 06056+2131 observational parameters.}
\label{tab:MERLIN_data}
\renewcommand{\arraystretch}{1.1}
\begin{tabular}{|c|c|c||c|}
\hline
Date of observation & 6,7, 8 and 9 Feb.2007 \\
No. antenna & six (Cm, Da, De, Kn, Pi and Mk2) \\
Field centre (RA, Dec J2000)&$\!\!$06$^h\!$ 08$^m\!$ 40$^s$.97, 21$\degree\!$ 31$^\prime\!$ $00\farcs60$ $\!\!$ \\
Rest frequencies (MHz)& 1665.402
1667.359 1720.530 \\
No. of frequency channels &  255 \\
Total band width (MHz) & 0.25 \\
Bandpass calibrator & 3C84 \\
Polarization angle calibrator & 3C 286 \\
Phase calibrator & 0617+210 \\
$\!\!\!\!$rms in quiet channel (Jy beam$^{-1}$)$\!\!\!\!$ & 0.03\\
\hline
\end{tabular}
      \end{center}
      \end{table*}
{We therefore measured the flux densities of the other polarization cubes at the same position as the error-weighted Stokes $I$ position. In the case of $Q$, one 2$\sigma_{\mathrm{rms}}$ result is given
because the other polarization products are significant.
We did this in two ways, firstly by using imfit to fit an unresolved Gaussian component the size of the restoring beam and secondly by using imstat to measure the maximum (or minimum, for $Q$, $U$ and $V$) flux density within the beam area. The results of both methods were the same to within the
noise-based positional error.}
\section{Results and Data analysis}
\label{sec:analys}
{We detected OH maser emission at the 1665-MHz main line toward IRAS 06056+2131. At the time of our MERLIN observations, the other main line at 1667 MHz and the satellite line at 1720 MHz are absent}.

{We detected total intensity (Stokes $I$) maser components in 9 successive channels (Table~\ref{tab:components}). The absolute position of the total intensity peak is RA = 06$^h$ 08$^m$ 40$^s$.6791, Dec = 21$\degree$ 31$^\prime$ $6\farcs929$ at $V_{\mathrm{LSR}}$= 10~km~s$^{-1}$. The positions of total intensity components in other channels were close to this, within the noise-based errors. Thus, we adopted the error-weighted centroid of the Stokes $I$ emission as common position for all channels and polarizations, of RA = 06$^h$ 08$^m$ 40$^s$.6775, Dec = 21$\degree$ 31$^\prime$ $6\farcs918$, standard deviations of 0\farcs021 in RA and 0\farcs017 in Dec.}

\begin{figure}

\begin{center}
  \includegraphics[width=1.0\columnwidth]{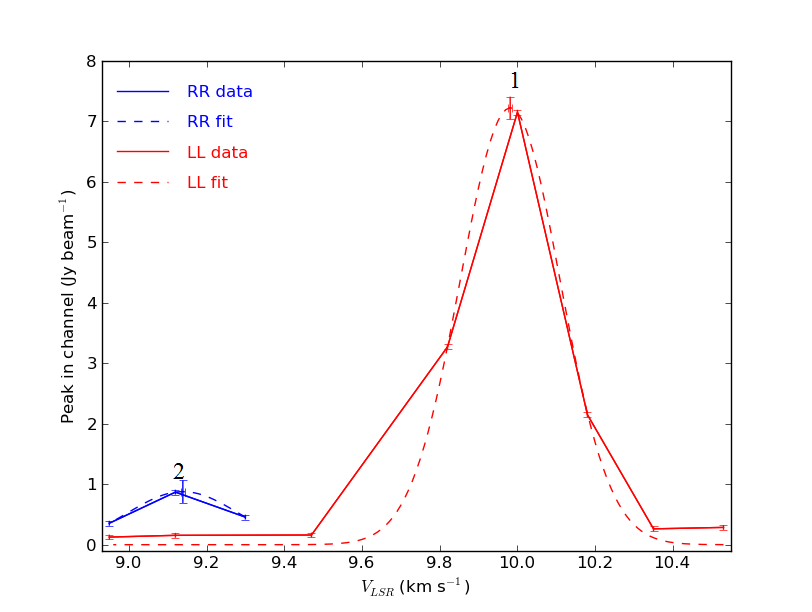}
  \caption{1665-MHz OH maser spectra towards IRAS 06056+2131 {at RA = 06$^h$ 08$^m$ 40$^s$.6775, Dec = 21$\degree$ 31$^\prime$ $6\farcs918$.} The LHC and RHC polarization peaks are labeled No.1 and No.2 respectively. Solid lines are used for the data while dashed lines are for the Gaussian fits.}
  \label{fig:I06056_RRLL}
\end{center}
\end{figure}

{Left-hand circular (LHC) polarization masers were detected in 8 of the 9 channels, and right-hand circular (RHC) in 4 channels. The LHC peak coincides with the total intensity peak. Their spectra are shown in Figure \ref{fig:I06056_RRLL}. If we assume that these are a Zeeman pair then all the OH 1665 MHz masers detected comprise a single feature.}

In order to measure the Zeeman splitting we fitted spectral Gaussian curves to the LHC and RHC peaks. The peak $V_{\mathrm{LSR}}$ and intensities are given in Table~\ref{tab:Zeeman}. The letter Z is associated with the RHC and LHC polarized peaks of the identified Zeeman pair. From the identified Zeeman pair, we are able to determine the line-of-sight
magnetic field ($B_{\parallel}$) in the massive star forming region IRAS 06056+2131 by measuring the velocity difference between the two polarization hands \cite{1996ApJ...457..415E}.

{We also measured the Stokes parameters ($I$, $Q$, $U$ and $V$),
following the radio definitions}:

\begin{align}\label{equ1}
I(\textit{u},\textit{v})&= 1/2\,[\mathrm{RR}(\textit{u},\textit{v})+\mathrm{LL}(\textit{u},\textit{v})]\\
V(\textit{u},\textit{v})&= 1/2\,[\mathrm{RR}(\textit{u},\textit{v})-\mathrm{LL}(\textit{u},\textit{v})]\\
Q(\textit{u},\textit{v})&= 1/2\,[\mathrm{RL}(\textit{u},\textit{v})+\mathrm{LR}(\textit{u},\textit{v})]\\
U(\textit{u},\textit{v})&= 1/2\textit{i}\,[\mathrm{LR}(\textit{u},\textit{v})-\mathrm{RL}(\textit{u},\textit{v})]
\end{align}%
$I$($\textit{u},\textit{v}$) is the Stokes integrated (total) flux density, $Q$($\textit{u},\textit{v}$) and $U$($\textit{u},\textit{v}$) are the Stokes flux densities corresponding to the two orthogonal components of linear polarization, while $V$($\textit{u},\textit{v}$) represents the circular component of polarized emission.
\begin{table*}
\caption{Measured peak flux densities per channel and errors ($\sigma_{\mathrm{rms}}$) in Stokes $I, Q, U, V$, $\mathrm{LL}$, $\mathrm{RR}$, polarized intensity $P$ and polarization angle $\chi$. The symbol -- means that no emission was detected above 3$\times$ the given $\sigma_{\mathrm{rms}}$ error. }
\begin{tabular}{crrrrrrrrrrrrrrrrr}
\hline
 $V_{\mathrm{LSR}}$&$I$&$I$err&$V$   & $V$err &$\mathrm{RR}$  & $\mathrm{RR}$err &$\mathrm{LL}$   & $\mathrm{LL}$err  &  $Q$  & $Q$err  & $U$ & $U$err & $P$  &  $P$err  &$\chi$  &$\chi$err\\
(km s$^{-1}$) &\multicolumn{2}{c}{(mJy b$^{-1}$)}&\multicolumn{2}{c}{(mJy b$^{-1}$)}&\multicolumn{2}{c}{(mJy b$^{-1}$)}&\multicolumn{2}{c}{(mJy b$^{-1}$)}&\multicolumn{2}{c}{(mJy b$^{-1}$)}&\multicolumn{2}{c}{(mJy b$^{-1}$)}&\multicolumn{2}{c}{(mJy b$^{-1}$)}&\multicolumn{2}{c}{(degrees)}\\
\hline
 10.53 & 131 &   19& --180 &  34&  --  &   40&  298 &   35 &  -- & 30 & --   &  30&--   & --   &--  &  --\\
 10.35 & 127 &   42& --197 &  42&  --  &   40&  282 &   65 &  -- & 30 & --   &  30& --  &  --  &--  &  --\\
 10.18 & 1119&   48& --1138&  51&  --  &   40&  2257&   89 &  -- & 30 & --   &  30& --  &  --  &--  &  --\\
 10.00 & 3808&  130& --3673& 133&  250 &   58&  7504&  250 &  112& 28 & 197  &  36&  217&    33&  29&    4\\
  9.82 & 1702&   60& --1665&  91&  --  &   40&  3375&  126 &   55& 30 & 203  &  31&  205&    34&  36&    4\\
  9.47 & 102 &   31&    -- &  30&  --  &   40&  138 &   55 &  -- & 30 & --   &  30& --  &  --  &--  &  --\\
  9.30 & 192 &   39&  299  &  29&  476 &   39&  --  &   40 &  -- & 30 & --   &  30& --  &  --  &--  &  --\\
 9.12 & 545 &   19&  391  &  35&  927 &   42&  164 &   34 &  -- & 30 & --   &  30& --  &  --  &--  &  --\\
  8.95 & 230 &   26&  163  &  38&  375 &   49&  251 &   58 &  -- & 30 & --   &  30& --  &  --  &--  &  --\\
\hline
\end{tabular}
\label{tab:components}
\end{table*}

\begin{table}
\caption{Gaussian fit parameters of the RHC and LHC polarization peaks of the 1665-MHz OH maser detected towards IRAS 06056+2131.}
\label{tab:zeeman}
\begin{center}
\renewcommand{\arraystretch}{1.1}
\begin{tabular}{|c|c|c|c|c|}
\hline
Peak  & $V_{\mathrm{LSR}}$  & Peak Flux &{Zeeman Pair}& $B_{\parallel}$\\
(km s$^{-1}$)&(Jy/beam)&(mG)\\
\hline
1. LHC   &9.981 &   7.675 & \multirow{2}{*} Z& \multirow{2}{*}{$-1.5$}\\
2. RHC   &9.139 &  0.934  & \\
\hline
\end{tabular}
      \end{center}
      \label{tab:Zeeman}
      \end{table}

{The degree of linear ($P_{\mathrm{linear}}$) and circular ($P_{\mathrm{circular}}$) polarization was calculated as}:

\begin{align}\label{equ5}
P_{\mathrm{linear}}&= (Q^{2}+U^{2})^{0.5}/I\\
P_{\mathrm{circular}}&= V/I
\end{align}
The total polarization degree $P_{\mathrm{total}}$ can be then calculated from Eq.7 as:

\begin{align}\label{equ7}
P_{\mathrm{total}}&= (Q^{2}+U^{2}+V^{2})^{0.5}/I
\end{align}
From Equations 3 and 4, the polarization position angle $\chi$ was calculated as:
\begin{align}\label{equ8}
\chi &= 0.5\times\arctan(U/Q)
\end{align}%
The Stokes and other polarization parameters were measured from the data cubes as described in Section~\ref{sec:obser} and listed in Table~\ref{tab:components}. Note that the POLI cube corresponds to $(Q^{2}+U^{2})$$^{0.5}$ and POLA to $\chi$.
Once we determine $P_{\mathrm{linear}}$ and $P_{\mathrm{circular}}$ polarization degrees from equations 5 and 6 respectively, the percentage of those parameters can be identified by \textit{m$_{\mathrm L}$}, \textit{m$_{\mathrm C}$} and \textit{m$_{\mathrm T}$} for the linear, circular and total polarization, respectively and are listed in Table~4.
\begin{table}
        \caption {The polarization angle ($\chi$) and percentages of linear (m$_{_{\textit L}}$), circular (m$_{_{\textit C}}$) and total (m$_{_{\textit T}}$) polarization, with respect to total intensity,  of the 1665-MHz OH maser
    emission detected towards IRAS 06056+2131.}
    \label{tab:poldegree}
        \begin{tabular}{p{15mm}p{15mm}p{15mm}p{15mm}}
    \hline
    $\chi$   & $m_{\mathrm{L}}$ & $m_{\mathrm{C}}$& $m_{\mathrm{T}}$\\
    ($\circ$)&       ($\%$)     &     ($\%$)       &   ($\%$)\\
    \hline
          29    &         6.0      &      96.5        &    96.6\\
    \hline
		\end{tabular}
\end{table}

\section{Discussion}
\label{sec:discu}
{Our {MERLIN} OH maser observations represent the first polarimetric study at high spectral and angular resolution of IRAS 06056+2131. The first detection of OH maser emission towards this source was made by \cite{1979A&AS...37....1T}. They observed the 4 ground-state OH transitions. Though they tabulated putative detections of 1667 MHz maser emission, the strongest component, at $V_{\mathrm{LSR}}$=9.2~km~s$^{-1}$, is in the range 3 to 5 sigma according to their peak-to-peak noise values. They nonetheless made a clear detection of the 1665-MHz transition, centred at a velocity of $V_{\mathrm{LSR}}$=8.9~km~s$^{-1}$ with a full width half maximum of 4.6~km~s$^{-1}$ (note that their channel resolution is 1.4~km~s$^{-1}$).}

{The detection of 1665-MHz OH emission, centred at $V_{\mathrm{LSR}}$=10~km~s$^{-1}$, was then confirmed by \cite{1988MNRAS.231..205C} within 1$^\prime$ from the IRAS source. They measured a total flux density of 3.4 Jy and 0.3 Jy in the LHC and RHC polarization respectively.}
{Moreover, \cite{2007A&A...465..865E} using Nan\c{c}ay and Green Bank Telescope observations detected emission at 1667 and 1720 MHz in addition to the 1665 MHz line at velocities 9.44, 3.37 and 10.14~km~s$^{-1}$ with flux densities 0.22, 0.6 and 3.23 Jy respectively.}
{Despite the similarity in velocities of the 1667 and 1665 MHz masers,
our non-detection at 1667 MHz could be due to variability (since it
was previously much weaker than 1665 MHz) but could also indicate that
the environment is becoming warmer (\citealt{2002MNRAS.331..521C};
\citealt{1991MNRAS.252...30G}).}

{Previous surveys of the 1720 MHz OH maser line by \cite{2003ApJ...596..328F} and \cite{2016ApJ...822..101R} suggested that this line is clearly observed from regions with relatively high magnetic field strengths which coincide with H\,{\small II} regions.}
{The significant difference in the velocity between the 1720-MHz line detected by \cite{2007A&A...465..865E} and the main lines} could be an indication that the 1720-MHz is tracing another protostellar object or that it is excited by different phenomena.
\subsection{Polarization}
\label{subsec:pol}
\begin{figure}
\hspace*{-0.5cm}
    \includegraphics[width=1.1\linewidth]{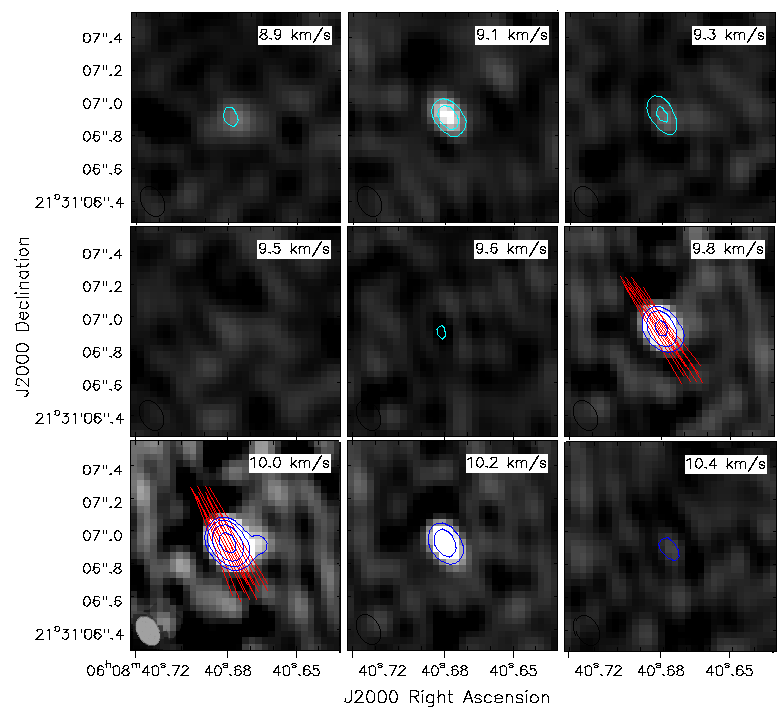}
\caption{Channel maps of IRAS 06056+2131. The contour levels are adapted to the dynamic range per panel. The grey scale shows the total intensity, cut at [-0.1,1] Jy. The light blue contours show Stokes $V$ at (1, 2) x 12 mJy/beam. The dark blue contours show Stokes $V$ at (1, 2, 4, 8) $\times$ --32 mJy/beam in the first panels, --12 mJy/beam ($3\sigma_{\mathrm {rms}}$) in the last panel (at $V_{\rm LSR}$=10.4~km~s$^{-1}$). The filled beam is shown at lower left of the panel at $V_{\rm LSR}$=10.0~km~s$^{-1}$. The red vectors represent the linear polarization position angle where the linear polarization is significant (at a threshold of $3\sigma_{\mathrm {rms}}$). See Table~\ref{tab:components} for details.}
    \label{fig:I06056_I-V-PA}
    \end{figure}

\begin{figure}
    \center
    \includegraphics[width=1.0\linewidth]{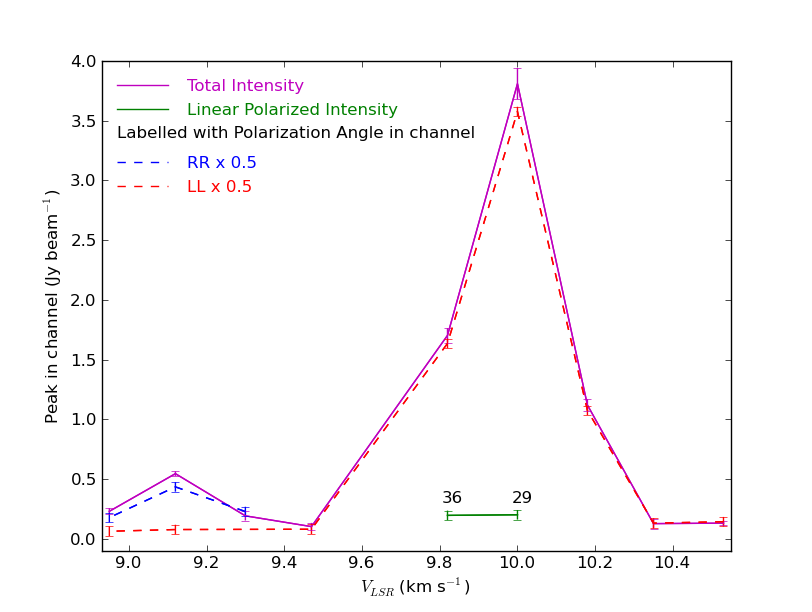}
    \caption{The total and polarized intensity spectra of 1665~MHz toward IRAS 06056+2131. The green solid bar is labeled with the linear polarization angle per channel. RR and LL have been multiplied by 0.5 to facilitate comparison with Stokes $I$ and provide a more compact plot.}
    \label{fig:I06056_I_P_PA}
\end{figure}

{The derived polarization parameters listed in Table 4 indicate that IRAS 06056+2131 is highly circularly polarized source. However, at the velocity of the LHC peak we found a linear polarization percentage of 6\% and a circular polarization percentage of 96.5\%, which means that the emission is elliptically polarized.
The line-of-sight component of the magnetic field at the location of the Zeeman pair is $B_{\parallel}= -1.5$~mG,  indicating a magnetic field pointing toward us.
{Figure \ref{fig:I06056_I-V-PA} shows the detected OH maser emission through different channels of IRAS 06056+2131 with their circular polarization contours and linear polarization vectors. Figure \ref{fig:I06056_I_P_PA} shows the total intensity (Stokes~$I$) of IRAS 06056+2131, where we can see clearly the Zeeman pair. It also displays the measured linear polarization intensity labeled with their polarization angles.}
{The contribution of the magnetic field in the plane of the sky, which is perpendicular to the polarisation vectors could suggest that it is broadly aligned with the possible SE-NW CO outflow detected by \cite{2006AJ....132...20X}, see Section \ref{sec:evolution} and Figure \ref{fig:I06056xy}}. However, the foreground Faraday rotation can potentially be strong at low frequencies hence making it harder to interpret the magnetic field direction \citep[e.g.][]{2006MNRAS.371L..26V}.

In order to estimate this effect on the orientation of the measured polarization position angle ($\chi$), we used the same method as in \citet{2010MNRAS.406.2218E} relying on the measurement of the dispersion measure (DM) and the rotation measure (RM) from pulsars. The closest pulsar to IRAS~06056+2131 is found to be PSR~J0609+2130. This pulsar is located at 1.2~kpc and has a dispersion measure (DM) of 38.77~cm$^{-3}$~pc \citep{2004MNRAS.347L..21L}. Unfortunately, no RM is provided for this pulsar. The next closest pulsar with both a known DM of 96.91~cm$^{-3}$~pc and a known RM of 66.0~rad~m$^2$, PSR B0611+22, is located with an offset of 1.6$^\circ$ from IRAS~06056+2131 at a comparable distance of 1.74~kpc (Sobey, priv. communication).

Assuming that the ambient Galactic magnetic field at the location of PSR~J0609+2130 and PSR~B0611+22 is similar, the RM for PSR~J0609+2130 is then inferred to be 26.4~rad~m$^2$, based on Eq.~1 from \cite{1999MNRAS.306..371H}, given by:
\begin{align}\label{equ9}
B_{\parallel}= 1.232 (\mathrm{RM/DM})
\end{align}}%
Since the degree of Faraday rotation is given by RM$\lambda$$^{2}$ \citep{2008MNRAS.386.1881N}, this implies that the values of $\chi$ measured here potentially suffer from Faraday rotation as high as 50$^\circ$.

Additionally, internal Faraday rotation can also affect the linear polarisation information if the OH masers are located right behind the UC\,H\,{\small II} region. Using \cite{2006ApJS..164...99F} eq.~5 and the electron density in addition to the diameter of the masing region from \citet{1994ApJS...91..659K} (cf. their Table 5), would lead to a very high value of internal Faraday rotation. On the other hand, \cite{2006ApJS..164...99F} show that even if there is a substantial amount of Faraday rotation (i.e., >1 rad) along the amplification length, the key length is the gain length. The fact that we measure a significant percentage of linear polarisation implies that internal Faraday rotation potentially affecting the measurements of $\chi$ along the gain length is < 1 rad (i.e., <57$^\circ$), suggesting that the masers arise from the limb or the near side of the UC\,H\,{\small II} region.
Since the sense of the external and internal potential Faraday rotation is nonetheless not known, the measurements of  $\chi$ can suffer from 0$^\circ$ to up to ~110$^\circ$ of rotation.

The alignment of the polarisation vectors with the orientation of the CO outflow  can consequently be fortuitous. It is to be noted though, that the velocity of the masers is in good agreement with their suggested association with the red lobe of the CO outflow.


{We performed a rough comparison between the energy density in the magnetic field of strength $B$ and the kinetic energy density in the CO outflow (see also Section~\ref{sec:evolution}) in order to investigate whether the magnetic field could have a significant effect, following the method summarised in \citet{2013MNRAS.431.1077A}. We assume that $B=B_{\parallel}$; if $B$ is greater then the magnetic energy density will be even greater. The magnetic energy density is given by $E_{\mathrm B} = B^2/(2 \mu_0)$ where $\mu_0$ is the permeability of free space.  For $B$ in Gauss, this gives $E_{\mathrm B} \approx 0.004 B^2 \approx 9\times10^{-9}$ J m$^{-3}$.}

{The thermal energy density is given by $E_{\mathrm{th}} = 3/2 n
k_{\mathrm B} T_{\mathrm K}$ where $k_{\mathrm B}$ is Boltzmann's
constant. The non-detection of the 1667 GHz maser implies that the
gas temperature $T_{\mathrm K} > 75$ K and the number density $n >
10^{10}$ m$^{-3}$ (\citealt{2002MNRAS.331..521C};
\citealt{1991MNRAS.252...30G}).  This leads to $E_{\mathrm{th}} > 1.5 \times
10^{-11}$ J m$^{-3}$.  Even for a much higher temperature of 200 K,
$E_{\mathrm{th}} < E_{\mathrm B}$ as long as $n < 2 \times 10^{12}$
  m$^{-3}$.}

{The bulk energy density is given by $E_{\mathrm{Bulk}} = 0.5 \rho v^2$ where we
assume the velocity $v$ as the CO outflow velocity of 7.5 km s$^{-1}$,
half the maximum velocity span measured by \cite{2006AJ....132...20X}.
The gas density $\rho = n \times 1.25 m_{\mathrm H_2}$, where
$m_{\mathrm H_2}$ is the mass of the hydrogen molecule. Thus,
$E_{\mathrm{Bulk}} < E_{\mathrm B}$ for $n < 8\times 10^{10}$ m$^{-3}$.}

These estimates are very crude but show that the magnetic field is
likely to influence the outflow for number densities in the range $1 - 8
\times10^{10}$ m$^{-3}$ ($1-8 \times10^{7}$ cm$^{-3}$) within a temperature range of $75 - 200$ K.
\subsection{The evolutionary status}
\label{sec:evolution}
\begin{figure}
  \centering
  \includegraphics[width = 1.0 \linewidth]{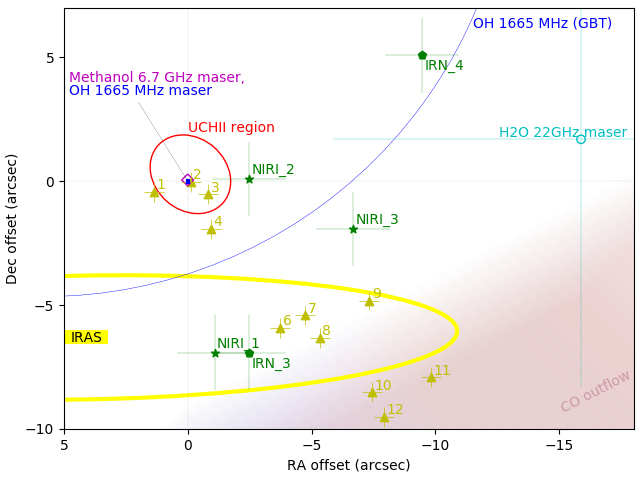}
  \caption{Overview of the star-forming region IRAS 06056+2131.
    The origin of coordinates (0,0) is the location of the OH maser peak at RA = 06$^h$ 08$^m$ 40$^s$.6775, Dec = 21$\degree$ 31$^\prime$ $06\farcs918$, shown by the blue cross. The error circle of the 1665-MHz observation by \protect\cite{2007A&A...465..865E} is shown by the blue arc. The magenta diamond and cyan circle mark the methanol \protect\citep{2009A&A...507.1117X} and water \protect\citep{1989A&A...221..295K} masers, respectively. The red ellipse marks the UC\,H\,{\small II} region \protect\citep{2016ApJ...833...18H}. The dark green star symbols mark the IR sources identified by \protect\cite{2018AJ....156....1K}, while the light green triangle symbols labeled from 1 to 12 represent the Smithsonian Millimetre Array (SMA) detections by \protect\cite{2012BAAA...55..199R}. The yellow ellipse marks the IRAS source position bounds, and the shaded area marks an estimate of the CO outflow direction from \protect\cite{2006AJ....132...20X}. See Section~\ref{sec:discu} for details.}
     \label{fig:I06056xy}
\end{figure}

{To understand the evolutionary status of the star forming region IRAS 06056+2131, we put together all the results of the different observations with known position performed close to the OH maser. This is presented in
Figure \ref{fig:I06056xy}. This figure has the MERLIN OH maser position,
 RA = 06$^h$ 08$^m$ 40$^s$.6775, Dec = 21$\degree$ 31$^\prime$ $6\farcs918$ at the origin, marked by a blue cross the size of which represents the position error ($0\farcs025$)}.

We estimated the positions of the OH masers observed using the
Greenbank Telescope (GBT) using maps provided by \citep{2007A&A...465..865E}. The 1665-MHz LHC
maser peak at 10.14~km~s$^{-1}$, is at RA = 06$^h$ 08$^m$ 41$^s$.095, Dec = 21$\degree$ 31$^\prime$ $21\farcs21$, uncertainty 38 arcsec. This is offset by (5.8, 14.3) arcsec from the
MERLIN position, within the GBT uncertainty.
{However, the 1720 MHz maser peak, at 3.4~km~s$^{-1}$, is at RA = 06$^h$ 08$^m$ 44$^s$.348, Dec = 21$\degree$ 31$^\prime$ $40\farcs01$, uncertainty 40 arcsec, an offset of (51, 33) arcsec.}
{The highest-resolution observation of the 6.7~GHz methanol maser was made by \cite{2009A&A...507.1117X} using MERLIN, with $\sim$30~mas astrometric accuracy (based on analysis of the original data) at a peak position of RA = 06$^h$ 08$^m$ 40$^s$.671, Dec = 21$\degree$ 31$^\prime$ $06\farcs89$, velocity 8.8~km~s$^{-1}$. This is marked by the magenta diamond ($\sim10$ times larger than the position error) in Figure \ref{fig:I06056xy}. In a recent observation, \cite{2016ApJ...833...18H} detected the maser at the same position within uncertainties, peaking at 9.39~km~s$^{-1}$}.

{\cite{1989A&A...221..295K}, using Effelsberg, found that the 22-GHz H$_{2}$O maser is offset by
--20$''$, 8$''$ from the IRAS source, position accuracy 10$\farcs$ , at a velocity 2~km~s$^{-1}$, shown by the cyan circle and error bars in Figure \ref{fig:I06056xy}.}

{\cite{2010A&A...517A..56F} using the Nobeyama 45-m telescope failed to detect any Class-I methanol maser emission lines (44 GHz and 95GHz) towards IRAS 06056+2131, On the other hand, their Effelsberg
observations of the Class-II 6.7 GHz methanol maser transition show a complex spectrum with emission in the entire velocity range ~[8.5;11]~km~s$^{-1}$, with at least 3 spectral components, the strongest being at 8.8~km~s$^{-1}$ with peak flux density = 12.7 Jy.}

{The most precise radio continuum observation of the UC\,H\,{\small II} region was made by \cite{2016ApJ...833...18H} using the VLA, giving a
position of RA = 06$^h$ 08$^m$ 40$^s$.67, Dec = 21$\degree$ 31$^\prime$ $07\farcs2$}. The astrometry is accurate to at least $0\farcs3$. The extent of the UC\,H\,{\small II} region is shown by the red ellipse (with angular size of 3$\farcs$5 x 2$\farcs$9 at position angle 41.4$\degree$) in Figure \ref{fig:I06056xy}.
{The position of IRAS~06056+2131 is RA = 06$^h$ 08$^m$ 40$^s$.973, Dec = 21$\degree$ 31$^\prime$ $00\farcs61$, with an error ellipse of axes (30, 5) arcsec with the long axis almost exactly E-W \citep{joint1988infrared}(shown in yellow in Figure \ref{fig:I06056xy}), is likely to represent an aggregate of sources within the IRAS beam. Several of the Infra-red
Reflection Nebulae (IRN) and Near Infra-Red Illuminators (NIRI)
identified by \cite{2018AJ....156....1K} (cf. their Tables 1 and 2) are within 10
arcsec of the OH maser. We assumed that the astrometric accuracy was
typical for the detector used, probably $\sim$ $1\farcs5$ ($\sim$3 pixels). These sources are shown and labeled in dark green in Figure \ref{fig:I06056xy}.
Note the existence of a WISE source, at RA = 06$^h$ 08$^m$ 40$^s$.45, Dec = 21$\degree$ 31$^\prime$ $02\farcs0$, which is within the positional uncertainty of the IRAS source \citep{2018AJ....156....1K}}. Consequently, as noted by
\cite{2018AJ....156....1K} the WISE source probably corresponds to the IRAS source and is not shown separately in Figure \ref{fig:I06056xy}.

{\cite{2012BAAA...55..199R} detected 12 mm-wave sources in the
region, using the SMA. We assumed that the published positions
(estimated from their figure 1) were relative to the pointing position
RA = 06$^h$ 08$^m$ 40$^s$.31, Dec = 21$\degree$ 31$^\prime$ $03\farcs6$ listed in the SMA (Smithsonian Millimetre Array) archive, with an astrometric accuracy of 0\farcs4 (1/3 of a synthesised beam). These sources are shown and numbered in light green in Figure \ref{fig:I06056xy}}.

{Figure \ref{fig:I06056xy} shows that the hot core SMA 2, the OH 1665 MHz
maser and the methanol maser are closely associated. SMA 2 is
approximately 0.130 arcsec west of the OH maser, within the position error.
The methanol maser is (0.095$\pm$0.039) arcsec from the OH maser, which  corresponds to $\sim140\pm60$ au in the plane of the sky at assumed distance of 1.5~kpc.
Such a small separation between the OH and Class II methanol masers indicates that they are
from a very similar region at overlapping velocities (around 8--10
~km~s$^{-1}$). These all lie in the direction of the
UC\,H\,{\small II} region, in fact SMA 2 is the closest hot core to the region
centre}.

{The close association (including similar velocities) between OH and Class II methanol masers is likely to indicate a relatively later evolutionary stage of massive star formation than regions which only have methanol masers \citep[e.g][]{1997MNRAS.289..203C,2010MNRAS.401.2219B}.}

{The water maser is offset from these masers by much more than the combined errors and
peaks at a lower velocity($\sim 2$~km~s$^{-1}$). The position and the velocity of the water maser suggests that it might be pumped by shocks associated with the CO outflow.}

{The pale red-blue shaded strip represents the orientation of the CO outflow mapped by \cite{2006AJ....132...20X}
from their figure 1, along the SE -- NW axis  joining the blue- and red-shifted peaks.
It should be noted that CO emission was detected throughout the area and the association of these peaks with the same outflow is tentative.}

{\cite{2005ApJ...625..864Z}, using the NRAO 12 m telescope detected a CO J=2-1 outflow which they associated with a deeply embedded IR source and the UC\,H\,{\small II} region. These are located at an offset of(--14\farcs0,14\farcs0) in RA and Dec from the IRAS source \citep[see figure 2 of][]{2005ApJ...625..864Z}. \cite{2006AJ....132...20X} carried out higher resolution imaging of CO J=2-1 using the Nobeyama Radio Telescope. The integrated emission peak is at 2.5~km~s$^{-1}$ and two pairs of red and blue-shifted peaks are resolved, covering a total velocity range of about - 5 to +10~km~s$^{-1}$. The CO outflow appears to be oriented close to the line of sight although \cite{2018AJ....156....1K} point out that at high resolution the region is complex and other interpretations are possible. The OH maser velocity appears to be close to the most red-shifted
CO velocity of 10 km~s$^{-1}$ and the polarization vectors
(which are perpendicular to the magnetic field lines),
suggest that the magnetic field orientation might be in agreement
with that of the SE-NW CO outflow reported by \cite{2006AJ....132...20X} (see their figure 1). The direction indicated by $B_{\parallel}$ suggests that the magnetic field associated with the far outflow
points towards the observer.}
{In addition to the outflow, \cite{2018ApJS..235...31Y} reported evidence for infalling material, using observations of lines of {HCO$^+$ (1$-$0)}, {HCO$^+$~(3$-$2)}, {H$_2$CO~(2$_{12}$$-$1$_{11}$)} and {H$^{13}$CO$^+$~(1$-$0)}.

{The presence of both outflow and infall material could be additional evidence for a protostellar object in an early evolutionary stage, when the radiation force is still lower than the gravitational force at the outer boundary.}

{The observations conducted toward IRAS 06056+2131 suggest that it is a high-mass star-forming region at a not so early evolutionary phase, associated with an UC\,H\,{\small II} region detected by \cite{1989A&A...221..295K,1994ApJS...91..659K} and at highest resolution by \cite{2016ApJ...833...18H}}.
\subsection{Comparison with other sources}
{Main-line OH masers towards the IRAS sources 20126+4104 and 19092+0841
  (hereafter IRAS 20126, 19092) were also imaged, using MERLIN, at a
  similar angular resolution to this study, by
  \cite{2005A&A...434..213E} and \cite{2007A&A...465..865E},
  respectively. The velocity range of the OH maser emission at 1665-MHz detected towards them is 17 and 3~km~s$^{-1}$, respectively.

IRAS 20126 is at 1.7 kpc \citep{1989ApJ...345..257W}, a similar
distance as our target IRAS 06065+2131, which
  has a velocity range of the maser emission at 1665-MHz of
  1.5~km~s$^{-1}$.} In IRAS 20126, the two brightest OH maser
  features at 1665 MHz (peak fluxes > 0.9Jy), which are separated by
  ~14~km~s$^{-1}$, are so bright that they would have been detected with
  MERLIN toward both IRAS 19092 (D=4.48~kpc,
  \citealt{1996A&A...308..573M}) and IRAS 06056+2131. However, these two sources show an
intrinsically smaller $V_{\mathrm{LSR}}$ span than IRAS 20126}.

Compared with IRAS sources 20126 and 19092 (with extents of 2000 and 22400 au, respectively), the OH maser is very compact in IRAS 06056+2131 with an upper limit to its distribution of 0.03 arcsec which corresponds to 45~au at an assumed distance of 1.5 kpc. Both OH maser main lines were detected toward IRAS 19092 while only 1665 MHz was found toward IRAS 20126 and IRAS 06056+2131, suggesting a higher gas temperature and density for the latter two sources \citep{1991MNRAS.252...30G}.

{There is no evidence for OH masers tracing a circumstellar disk either in IRAS 06056+2131 or IRAS 19092, while \cite{2005A&A...434..213E} reported that the OH masers are tracing a circumstellar disk in IRAS 20126. The small upper limit to the size of the OH maser towards IRAS 06056+2131 makes it improbable that the small observed $V_{\mathrm{LSR}}$ range is a projection effect of an outflow in the plane of the sky.}

{IRAS 06056+2131 is relatively less luminous in comparison with IRAS 20126 and IRAS 19092 (5.83 $\times10$ $^{3}$ L$_{\odot}$, 10$^{4}$ L$_{\odot}$ and 10$^{4}$ L$_{\odot}$, respectively) while the line-of-sight magnetic field measured from the OH Zeeman pair has the lowest magnitude (1.5, 11 and 4.4 mG, respectively). The 6.7 GHz methanol (Class II) maser has been detected toward the three IRAS sources, while class I has been detected only toward IRAS 19092.}

{No radio-continuum was detected towards IRAS 20126 \citep{1998A&A...336..339M} and \citet{2017A&A...608A..80E} showed that the nearest detection is offset by about 2 arcmin from IRAS 19092, suggesting that the OH masers in these sources are not associated with UC\,H\,{\small II} regions, unlike  the presence of a UC\,H\,{\small II} region associated with IRAS 06056+2131.}

{Bearing in mind all the results summarised above, we suggest that IRAS 06056+2131 is at an evolutionary stage comparable with IRAS 20126, which is more evolved than IRAS 19092. This result is consistent with what was reported by \citet{2007A&A...465..865E}, where they found the sources with higher OH intensity to be more evolved. The absence of any Class I methanol maser from IRAS 06056+2131 is a further evidence that it is more evolved than IRAS 19092 \citep{2006ApJ...638..241E}.}

\section{Conclusions}
\label{sec:concl}
{We have presented a high angular resolution observations of OH maser emission toward IRAS 0605+2131. At the time of the observation of the three OH transitions at 1665, 1667 and 1720 MHz, only 1665 MHz was detected.
The small upper limit to the size of the OH maser emitting region is estimated to be $\sim$ 45~au at 1.5~kpc.
We measured the line-of-sight magnetic field from the identified
OH Zeeman pair to be $\sim$ --1.5 mG, and the corresponsing magnetic field energy density is strong enough to influence the outflow. The linear polarization vectors might suggest that the magnetic field orientation in the plane of the sky is roughly NW to SE. This might then be aligned with the possible CO outflow direction \citep{2006AJ....132...20X}. However, Faraday rotation gives an uncertainty of 110$^\circ$ in the actual magnetic field direction}.

{Our results are found to be in a good agreement with \cite{1998MNRAS.297..215C}, which tested the association between OH and 6.7-GHz masers with a large sample and concluded that, 80$\%$ of OH masers have 6.7-GHz methanol maser counterparts.}
{The spatial and velocity coincidence between OH and Class II methanol maser indicates that they are probably tracing the same physical phenomena in IRAS 06056+2131, however there is no evidence of a circumstellar disk.}
{The close association of a Class-II methanol maser and UC\,H\,{\small
II} region with the OH 1665-MHz maser and the absence of Class-I methanol masers} suggest that IRAS 06056+2131 is
at a later stage of evolution than sources without detectable OH
masers. In comparison with other OH maser sources investigated by
\cite{2005A&A...434..213E} and \cite{2017A&A...608A..80E}, the
properties of IRAS 06056+2131 also suggests that is at a relatively
more evolved evolutionary stage.

{Finally, higher angular resolution would provide a better estimate of
the location and properties of the protostellar object driving the
IRAS 06056+2131 outflow, for example VLBI measurements of masers and
ALMA measurements of the mm/sub-mm core.}

\section*{acknowledgements}
{We gratefully thank Prof. R. Battye, M. Gray and R. Beswick, and the rest of the e-MERLIN team for guidance in reducing these data. We also remember the important role of the late Dr Jim Cohen in initiating this project. We thank the anonymous referee for very insightful and helpful comments which have improved this paper.
e-MERLIN is the UK radio interferometer array, operated by the University of Manchester on behalf of STFC.
We acknowledge the use of MERLIN archival data as well as NASA's Astrophysics Data System Service.
M.Darwish would like to acknowledge the Science and Technology Development Fund (STDF) N5217, Academy of Scientific Research and Technology (ASRT), Cairo, Egypt and Kottamia Center of Scientific Excellence for Astronomy and Space Sciences (KCSEASSc), National Research Institute of Astronomy and Geophysics (NRIAG). Our sincere thanks to S. Ellingsen and J. Allotey for their helpful discussion and to C. Sobey for pulsar data.}
\section*{Data availability}
The raw visibility data and the clean images cubes in FITS format for all OH maser polarization products of the MERLIN observations of IRAS 06056+2131 can be found at:
\url{https://doi.org/10.5281/zenodo.3961902}.
\bibliographystyle{mnras}
\bibliography{ref}
%
 \bsp	
\label{lastpage}
\end{document}